\documentclass[12pt]{article}
\usepackage{epsfig}
\usepackage{cite}

\begin{document}

\title{Forthcoming occultations of astrometric radio sources by planets%
\footnote{In: D. Behrend, K. D. Baver (Eds.), IVS 2010 General Meeting Proc., NASA/CP-2010-215864, 2010, 320--324.}}
\author{Victor L'vov, Zinovy Malkin, Svetlana Tsekmeister \\ Pulkovo Observatory, St. Petersburg, Russia}
\date{\vspace{-10mm}}
\maketitle

\begin{abstract}
Astrometric observations of the radio source occultations by solar system bodies may be of large interest
for testing gravity theories, dynamical astronomy, and planetary physics.
In this paper, we present an updated list of the occultations of astrometric radio sources by planets expected in the nearest years.
Such events, like the solar eclipses, generally speaking, can be only observed in a limited region.
The map of the shadow path is provided for the events occurred in regions with several VLBI stations and
hence the most interesting for radio astronomy experiments.
\end{abstract}

\section{Introduction}

Observations of occultations of compact radio sources by Solar System planets may be interesting for several astronomical and physical applications,
such as testing GR~\cite{Malkin2009}, improvement of planet orbits and their tie to ICRF~\cite{Linfield1992}, and planetary
researches~\cite{Leblanc1992,Black2000}.

Our previous computations of occultations of astrometric radio sources by planets and their close approaches were published in~\cite{Malkin2009}.
In this paper we present the updated list of the forthcoming occultations that may be interesting for radio astronomy observations.
The main differences with the previous work are the use of an extended astrometric source list, and computation of event maps to help better
planning of observations.

\section{Forthcoming occultations}

Most computations of the circumstances of occultations of geodetic radio sources by planets were performed using
the codes APPROACH and OCCULT, which utilize the Ephemeride Package for Objects of the Solar System (EPOS)
data and environment\footnote{http://neopage.pochta.ru/ENG/ESUPP/main.htm}.
Source coordinates were taken from the catalog of astrometric radio source positions of Leonid Petrov,
version 2009c\footnote{http://astrogeo.org/vlbi/solutions/2009c\_astro/}

The list of occultations is presented in Table~\ref{tab:occulttaions} with their basic circumstances.
One can see that most of the events are visible in regions with radio astronomy observatories, and several of them can be observed on many antennas.
The nearest most interesting event is the occultation of the source 1946--200 by Mars in February 2011 visible in North America with VLBA, VLA, GBT
and other radio astronomy facilities.

\begin{figure}

\parbox{0.45\textwidth}{
\begin{center}
\epsfclipon \epsfxsize=0.33\textwidth \epsffile{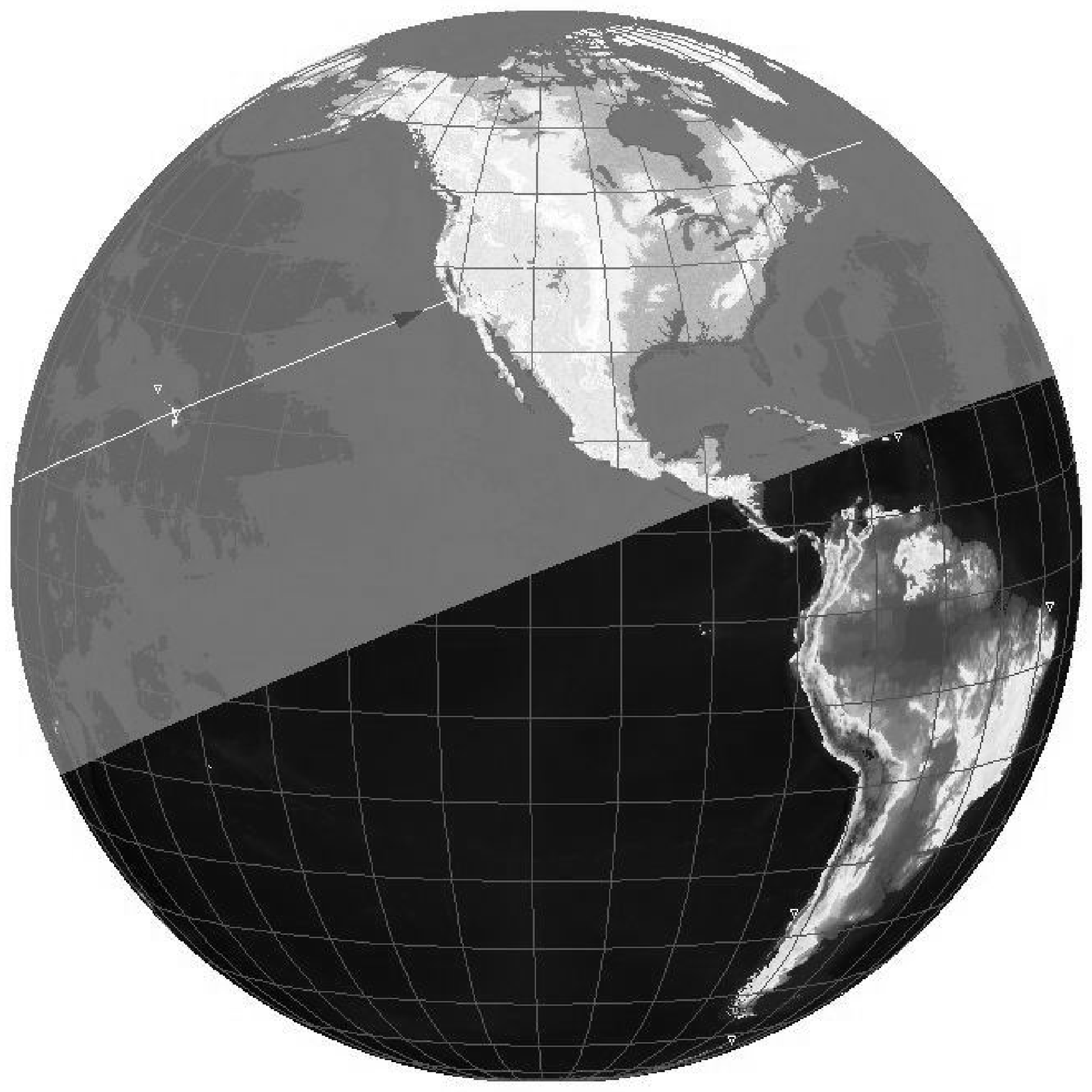} \\
0127+084, Mars, May 03, 2011
\end{center} }
\parbox{0.45\textwidth}{
\begin{center}
\epsfclipon \epsfxsize=0.33\textwidth \epsffile{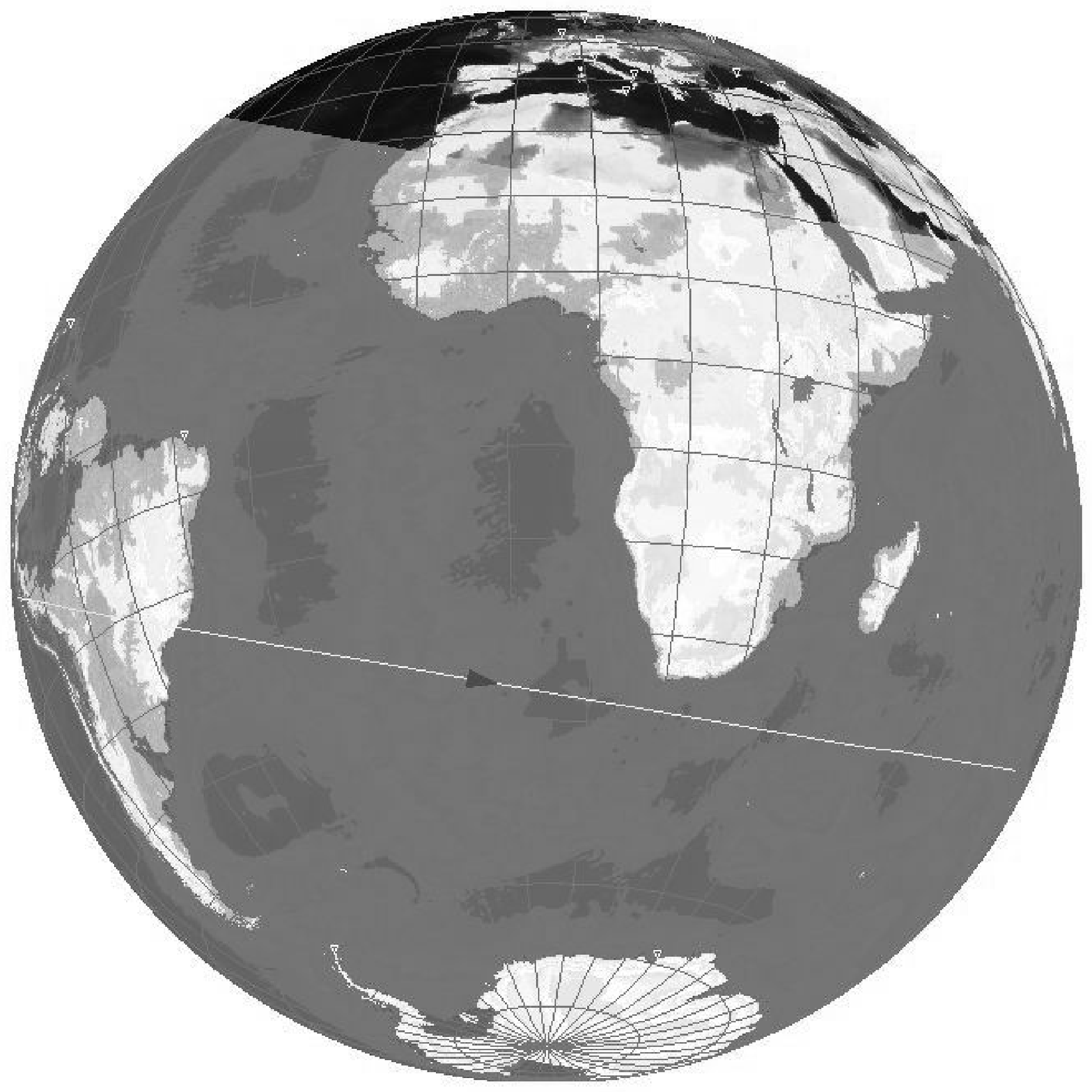} \\
1631--208, Venus, Dec 24, 2012
\end{center} }

\bigskip
\parbox{0.45\textwidth}{
\begin{center}
\epsfclipon \epsfxsize=0.33\textwidth \epsffile{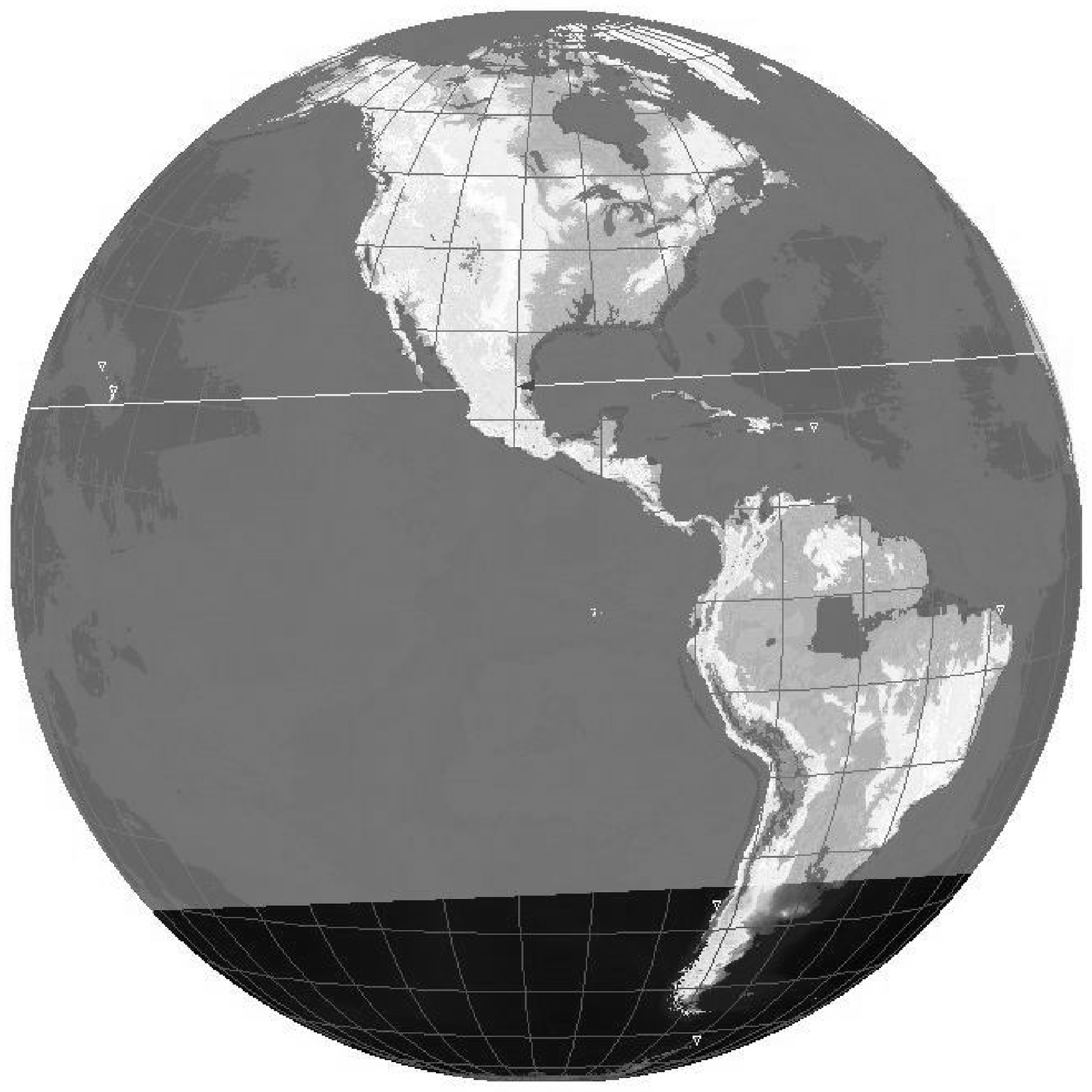} \\
0947+064, Venus, Aug 06, 2015
\end{center} }
\parbox{0.45\textwidth}{
\begin{center}
\epsfclipon \epsfxsize=0.33\textwidth \epsffile{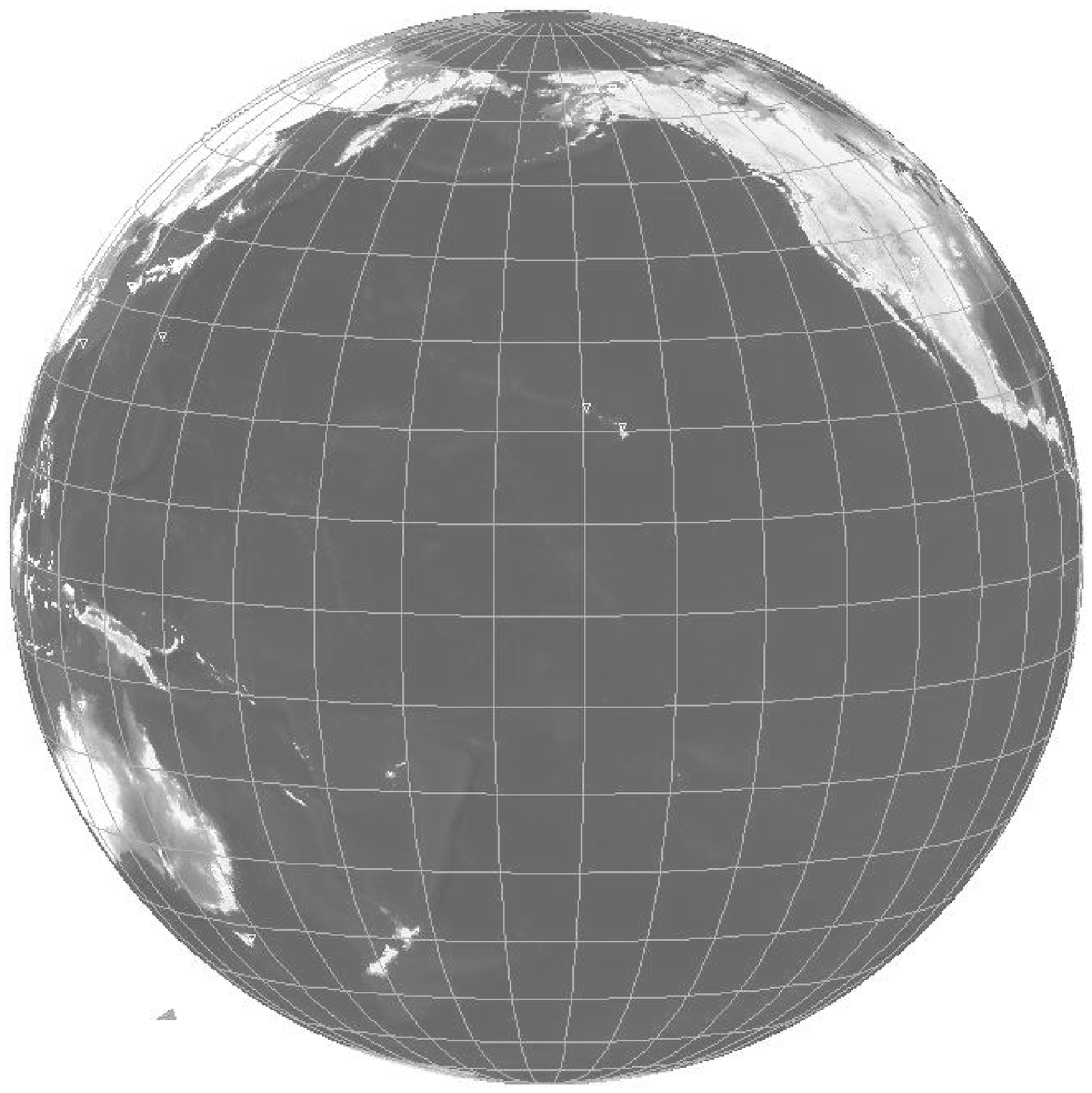} \\
1101+077, Jupiter, Apr 10, 2016
\end{center} }

\bigskip
\parbox{0.45\textwidth}{
\begin{center}
\epsfclipon \epsfxsize=0.33\textwidth \epsffile{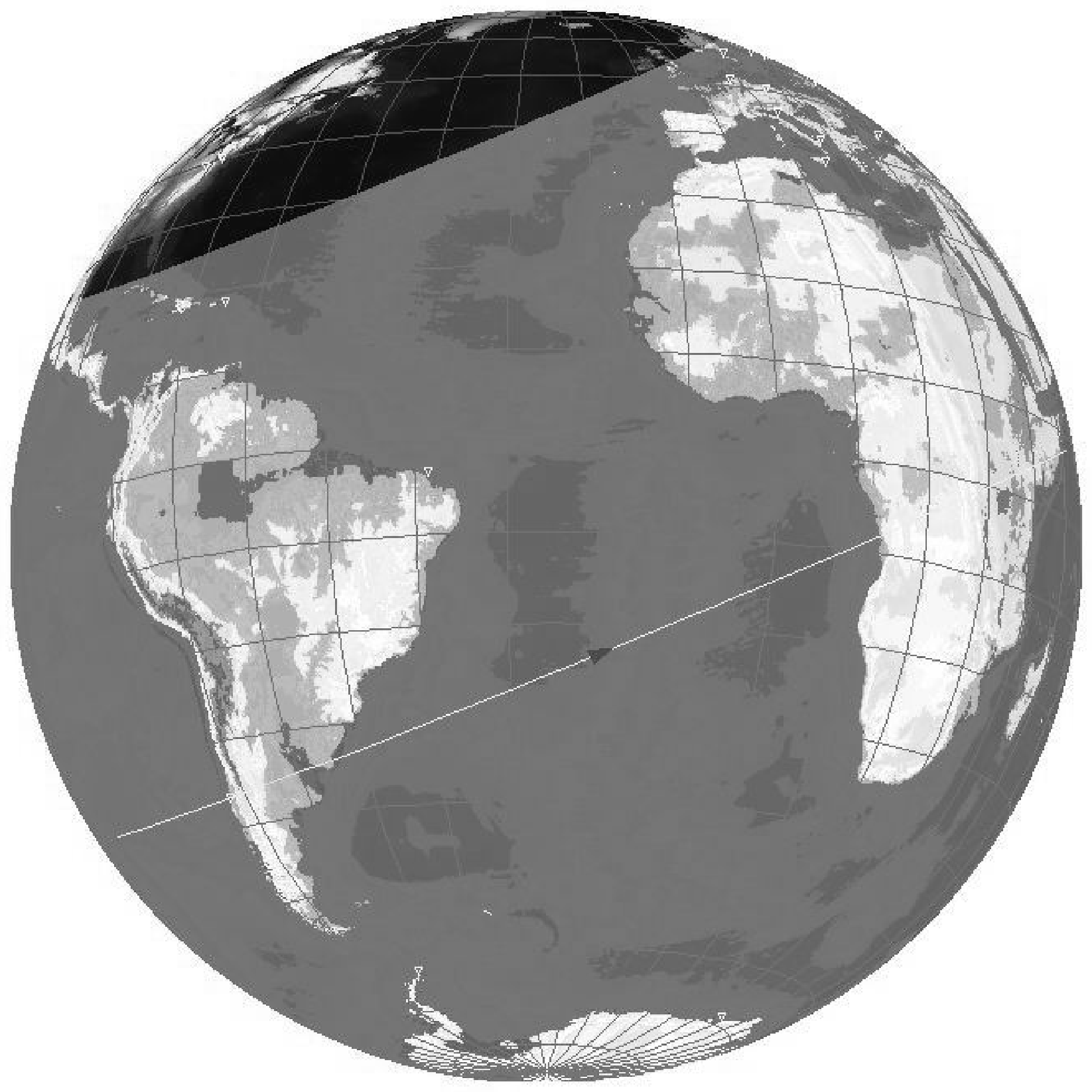} \\
2220--119, Venus, Jan 16, 2020
\end{center} }
\parbox{0.45\textwidth}{
\begin{center}
\epsfclipon \epsfxsize=0.33\textwidth \epsffile{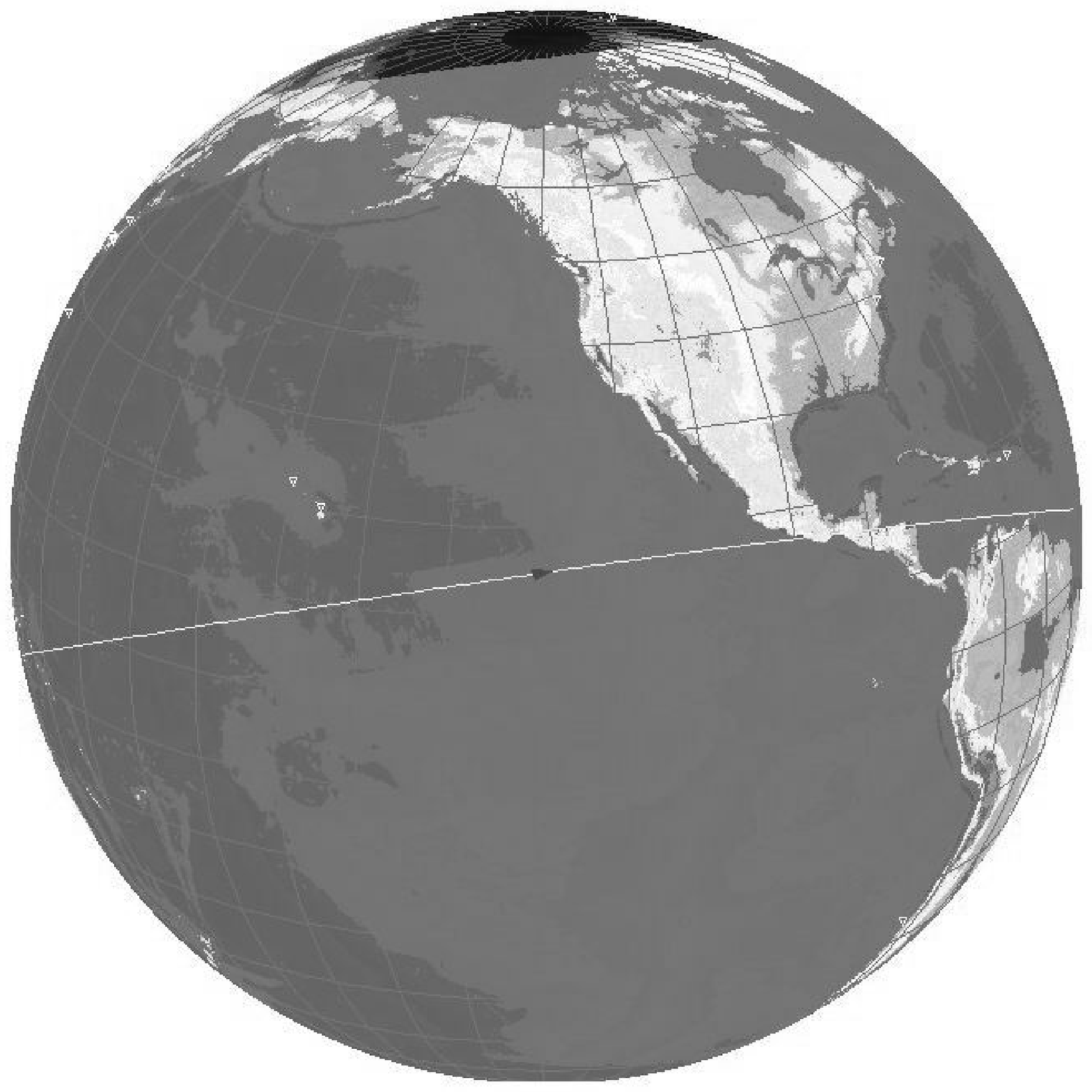} \\
0446+178, Venus, Jul 17, 2020
\end{center} }

\caption{Visibility of selected nearest occultations of radio sources by planets.
The region of visibility is shown in light (for the 2016 event, the shadow covers the whole hemisphere shown in the picture). }
\label{fig:maps}
\end{figure}

Figure~\ref{fig:maps} and Table~\ref{tab:details1} present the detailed circumstances of the several nearest events that can be observed
in regions with several geodetic VLBI antennas.
The maps of shadow path are shown in Fig.~\ref{fig:maps}. Table~\ref{tab:details1} presents the detailed circumstances of the several nearest events,
such as the elevation, azimuth, and position angle on the planetary limb at the beginning and the end of the occultation.
The circumstances for other events are available on request.

\begin{table}
\centering
\caption{Occultations of radio sources by planets in 2011--2030
(d is the angular distance from the Sun, the letter means east or west elongation).}
\label{tab:occulttaions}
\begin{tabular}{|l|ccc|c|r|l|}
\hline
Planet & \multicolumn{3}{|c|}{Date} & Source & d, deg & \multicolumn{1}{|c|}{Region of visibility} \\
       & Y & M & D                  &        &        & \\
\hline
Venus   & 2011 & 02 & 26 & 1946--200 &  42W  & Antarctic, S. America \\
Mars    & 2011 & 05 & 03 & 0127+084  &  19W  & N. America \\
Venus   & 2012 & 12 & 24 & 1631--208 &  23W  & S. America, Antarctic, Africa \\
Venus   & 2015 & 08 & 06 & 0947+064  &  15E~ & America \\
Jupiter & 2016 & 04 & 10 & 1101+077  & 144E~ & Australia, SE Asia \\
Venus   & 2020 & 01 & 16 & 2220--119 &  38E~ & S. America, Europe, Africa \\
Venus   & 2020 & 07 & 17 & 0446+178  &  42W  & N. America \\
Mercury & 2022 & 11 & 14 & 1529--195 &   4E~ & S. America \\
Jupiter & 2025 & 09 & 18 & 0725+219  &  65W  & America \\
Mercury & 2027 & 03 & 21 & 2220--119 &  27W  & N. America \\
Saturn  & 2028 & 10 & 24 & 0223+113  & 173W  & by ring; Asia, Europe, N. Africa \\
Mercury & 2029 & 01 & 14 & 1958--179 &   5E~ & Australia, Antarctic, S. Africa \\
Venus   & 2029 & 02 & 28 & 2221--116 &   6W  & Africa, SE Asia, Australia \\
Mercury & 2029 & 04 & 16 & 0243+181  &  19E~ & Asia, N. America \\
Mercury & 2029 & 12 & 27 & 1858--212 &   8E~ & S. America, Australia \\
Mercury & 2030 & 02 & 27 & 2208--137 &   9W  & S. America, Africa \\
\hline
\end{tabular}
\end{table}

\section{Conclusion}

Observations of the occultations of radio sources by planets are attractive for several interesting applications in physics and planetary science.
They can effectively supplement the observations of radio source occultation by the Moon, and spacecraft radio occultations by planets.
The list of occultations presented in this paper can be used for scheduling observations in different modes such as VLBI, connected-element
interferometer, or single-dish mode, depending on the scientific task.

The list of occultations as well as updated list of close approaches of planets to radio sources is available at
\verb"http://www.gao.spb.ru/english/as/ac_vlbi/".


\begin{table}
\centering
\caption{Detailed circumstances of the nearest occultations of radio source by planets: B~--- at the beginning, E~--- at the end.}
\label{tab:details1}
\tabcolsep=5.7pt
\begin{tabular}{|l|c|l|cc|rr|rr|rr|}
\hline
Planet & Date & Station & \multicolumn{2}{|c|}{Time, TT} & \multicolumn{2}{|c|}{El., deg} & \multicolumn{2}{|c|}{Az., deg} & \multicolumn{2}{|c|}{Pos. angle} \\
       &      &         & B  & E                         & B~ & E~                        & B~ & E~                        & B~ & E~ \\
\hline
Venus   & Feb 26, 2011 & OHIGGINS & 14:18:18 & 14:22:03 & 46 & 45 &  20 &  21 & 35 & 309 \\
        &              & SYOWA    & 14:18:20 & 14:23:11 & 11 & 11 & 118 & 119 & 55 & 290 \\
\hline
Mars    & May 03, 2011 & KOKEE    & 18:10:19 & 18:12:26 & 41 & 42 & 278 & 278 & 60 & 256 \\
        &              & MK-VLBA  & 18:10:19 & 18:12:28 & 46 & 46 & 278 & 278 & 66 & 250 \\
        &              & OV-VLBA  & 18:11:31 & 18:13:40 & 61 & 61 & 345 & 347 & 65 & 251 \\
        &              & DSS13    & 18:11:32 & 18:13:41 & 63 & 63 & 348 & 349 & 69 & 247 \\
        &              & KP-VLBA  & 18:11:38 & 18:13:45 & 67 & 67 & 359 & 360 & 77 & 239 \\
        &              & BR-VLBA  & 18:11:43 & 18:13:45 & 50 & 50 & 347 & 348 & 49 & 266 \\
        &              & PIETOWN  & 18:11:46 & 18:13:53 & 64 & 64 & 7   & 8   & 75 & 240 \\
        &              & LA-VLBA  & 18:11:50 & 18:13:57 & 63 & 63 & 11  & 12  & 74 & 241 \\
        &              & FD-VLBA  & 18:11:52 & 18:13:55 & 67 & 67 & 19  & 20  & 84 & 232 \\
        &              & GILCREEK & 18:11:59 & 18:13:20 & 28 & 29 & 318 & 319 & 17 & 298 \\
        &              & GGAO7108 & 18:12:37 & 18:14:41 & 47 & 46 & 55  & 56  & 84 & 232 \\
        &              & HN-VLBA  & 18:12:42 & 18:14:48 & 42 & 41 & 57  & 58  & 78 & 237 \\
        &              & WESTFORD & 18:12:43 & 18:14:49 & 41 & 41 & 58  & 58  & 79 & 237 \\
\hline
Venus   & Dec 24, 2012 & FORTLEZA & 10:06:07 & 10:09:25 & 45 & 46 & 244 & 244 & 80  & 300 \\
        &              & OHIGGINS & 10:07:13 & 10:10:03 & 31 & 32 & 286 & 287 & 136 & 244 \\
        &              & HARTRAO  & 10:08:02 & 10:11:28 & 68 & 67 &  83 &  83 & 87  & 293 \\
        &              & SYOWA    & 10:08:11 & 10:11:16 & 37 & 37 &  43 &  44 & 129 & 252 \\
\hline
Venus   & Aug 06, 2015 & YEBES40M & 18:41:27 & 19:23:00 &  9 & 1  &  91 &  98 & 284 &  68 \\
        &              & FORTLEZA & 18:44:31 & 19:20:16 & 39 & 31 &  79 &  81 & 236 & 117 \\
        &              & SC-VLBA  & 18:49:35 & 19:29:43 & 63 & 54 &  68 &  76 & 260 &  94 \\
        &              & WESTFORD & 18:53:34 & 19:32:45 & 50 & 46 &  30 &  43 & 284 &  70 \\
        &              & HN-VLBA  & 18:53:44 & 19:32:53 & 50 & 46 &  29 &  42 & 284 &  69 \\
        &              & GGAO7108 & 18:54:33 & 19:34:07 & 55 & 51 &  24 &  39 & 281 &  73 \\
        &              & FD-VLBA  & 19:02:27 & 19:42:29 & 63 & 66 & 334 & 357 & 274 &  80 \\
        &              & LA-VLBA  & 19:03:07 & 19:42:45 & 58 & 60 & 334 & 353 & 279 &  75 \\
        &              & PIETOWN  & 19:03:41 & 19:43:28 & 59 & 62 & 330 & 349 & 278 &  76 \\
        &              & KP-VLBA  & 19:04:48 & 19:44:46 & 59 & 63 & 322 & 341 & 276 &  78 \\
        &              & DSS13    & 19:06:13 & 19:45:56 & 54 & 59 & 318 & 334 & 279 &  75 \\
        &              & OV-VLBA  & 19:06:34 & 19:46:03 & 52 & 56 & 318 & 333 & 281 &  73 \\
        &              & BR-VLBA  & 19 06:39 & 19:44:22 & 43 & 46 & 324 & 336 & 291 &  63 \\
        &              & GILCREEK & 19:10:09 & 19:43:42 & 21 & 23 & 302 & 311 & 305 &  49 \\
        &              & MK-VLBA  & 19:17:05 & 19:58:37 & 31 & 41 & 275 & 279 & 269 &  86 \\
        &              & KOKEE    & 19:17:31 & 19:59:12 & 27 & 37 & 274 & 279 & 271 &  83 \\
\hline
\end{tabular}
\end{table}

\begin{table}
\centering
\addtocounter{table}{-1}
\caption{(continued)}
\label{tab:details2}
\tabcolsep=5.7pt
\begin{tabular}{|l|c|l|cc|rr|rr|rr|}
\hline
Planet & Date & Station & \multicolumn{2}{|c|}{Time, TT} & \multicolumn{2}{|c|}{El., deg} & \multicolumn{2}{|c|}{Az., deg} & \multicolumn{2}{|c|}{Pos. angle} \\
       &      &         & B & E                          & B~ & E~                        & B~ & E~                        & B~ & E~ \\
\hline
Jupiter & Apr 10, 2016 & TIGOCONC & 07:02:05 & 10:07:53 & 15 &-21 &  68 &  97 & 268 & 133 \\
        &              & LA-VLBA  & 07:05:34 & 10:15:00 & 49 & 13 &  54 &  90 & 274 & 128 \\
        &              & KP-VLBA  & 07:06:09 & 10:14:31 & 55 & 17 &  52 &  88 & 273 & 128 \\
        &              & DSS13    & 07:06:55 & 10:15:16 & 56 & 21 &  42 &  84 & 273 & 128 \\
        &              & OV-VLBA  & 07:07:09 & 10:15:37 & 55 & 22 &  38 &  83 & 273 & 128 \\
        &              & BR-VLBA  & 07:07:37 & 10:17:21 & 46 & 21 &  29 &  78 & 274 & 128 \\
        &              & GILCREEK & 07:10:03 & 10:20:51 & 32 & 26 & 351 &  45 & 274 & 127 \\
        &              & MK-VLBA  & 07:12:22 & 10:16:11 & 71 & 58 & 308 &  72 & 271 & 130 \\
        &              & KOKEE    & 07:12:53 & 10:16:52 & 67 & 60 & 306 &  65 & 271 & 130 \\
        &              & USSURISK & 07:15:37 & 10:25:14 &  7 & 40 & 267 & 305 & 273 & 130 \\
        &              & HOBART26 & 07:16:01 & 10:15:13 &  9 & 34 & 288 & 328 & 265 & 137 \\
        &              & VERAMZSW & 07:16:13 & 10:24:24 & 15 & 48 & 272 & 311 & 272 & 130 \\
        &              & TSUKUB32 & 07:16:27 & 10:24:25 & 14 & 50 & 271 & 308 & 272 & 131 \\
        &              & KASHIM34 & 07:16:28 & 10:24:21 & 14 & 50 & 271 & 308 & 272 & 131 \\
        &              & GIFU11   & 07:16:31 & 10:24:39 & 11 & 48 & 269 & 303 & 272 & 131 \\
        &              & SESHAN25 & 07:16:34 & 10:25:38 & -2 & 38 & 260 & 286 & 272 & 131 \\
        &              & TIDBIN64 & 07:16:41 & 10:16:04 & 12 & 41 & 288 & 327 & 266 & 136 \\
        &              & AIRA     & 07:16:48 & 10:24:57 &  6 & 45 & 265 & 294 & 271 & 131 \\
        &              & VERAIRIK & 07:16:48 & 10:24:59 &  6 & 45 & 265 & 294 & 271 & 131 \\
        &              & PARKES   & 07:16:56 & 10:16:27 & 12 & 43 & 287 & 325 & 266 & 136 \\
        &              & VERAISGK & 07:17:13 & 10:25:02 &  0 & 42 & 262 & 283 & 271 & 132 \\
        &              & CHICHI10 & 07:17:14 & 10:23:43 & 16 & 56 & 270 & 300 & 271 & 131 \\
        &              & VERAOGSW & 07:17:14 & 10:23:44 & 16 & 56 & 270 & 300 & 271 & 131 \\
\hline
Venus   & Jan 16, 2020 &          &   :  :   &   :  :   &    &    &     &     &     &     \\
\hline
Venus   & Jul 17, 2020 &          &   :  :   &   :  :   &    &    &     &     &     &     \\
\hline
\end{tabular}
\end{table}

\end{document}